\documentclass[twocolumn,preprintnumbers,amssymb]{revtex4}
\usepackage{graphicx}
\usepackage{dcolumn}
\usepackage{bm}
\usepackage{amssymb}
\usepackage{amsmath}
\begin{document}
\addtolength{\voffset}{-.75cm}
\addtolength{\textheight}{1.5cm} \addtolength{\hoffset}{-0.25cm}
\addtolength{\textwidth}{.5cm}
\title{Chameleon Vector Bosons}
\author{Ann E. Nelson$^{1,2}$ and Jonathan Walsh$^1$}
\affiliation{
$^1$Department of Physics, Box 1560, University of Washington,\\
           Seattle, WA 98195-1560, USA\\
           and\\
           $^2$Instituto de F\'isica Te\'orica UAM/CSIC\\
           Facultad de Ciencias, C-XVI\\
           Universidad Aut\'onoma de Madrid\\
           Cantoblanco, Madrid 28049, SPAIN
}

\newcommand{\geff}{g_{\rm eff}\ }
\newcommand{\meff}{m_{\rm eff}\ }
\newcommand{\lv}{\ell^{(0)}_V}
\newcommand{\lsm}{\ell_S}
\newcommand{\lvm}{\ell_V}
\newcommand{\ls}{\ell^{(0)}_S}
\newcommand{\p}[0]{\partial}
\renewcommand{\d}[0]{\textrm{ d}}
\newcommand{\R}[0]{\mathbb{R}}
\newcommand{\C}[0]{\mathbb{C}}
\newcommand{\bra}[1]{\big<#1\big|}
\newcommand{\ket}[1]{\big|#1\big>}
\renewcommand{\matrix}[4]{\left( \begin{array}{c c} #1 & #2 \\ #3 & #4 \end{array} \right)}
\newcommand{\pf}[2]{\frac{\partial #1}{\partial #2}}
\newcommand{\df}[2]{\frac{\textrm{d}#1}{\textrm{d}#2}}
\newcommand{\be}[0]{\begin{equation*}}
\newcommand{\ee}[0]{\end{equation*}}
\newcommand{\sfrac}[2]{\textstyle{\frac{#1}{#2}}}
\newcommand{\Res}[0]{\textrm{Res}}
\newcommand{\lc}[1]{\epsilon_{#1}}
\newcommand{\del}[0]{\nabla}
\newcommand{\braket}[2]{\big<#1\big|#2\big>}
\newcommand{\beq}{\begin{eqnarray}}
\newcommand{\eeq}{\end{eqnarray}}
\newcommand{\nn}{\nonumber}
\def\ltap{\ \raise.3ex\hbox{$<$\kern-.75em\lower1ex\hbox{$\sim$}}\ }
\def\gtap{\ \raise.3ex\hbox{$>$\kern-.75em\lower1ex\hbox{$\sim$}}\ }
\def\CO{{\cal O}}
\def\CL{{\cal L}}
\def\CM{{\cal M}}
\def\tr{{\rm\ Tr}}
\def\CO{{\cal O}}
\def\CL{{\cal L}}
\def\CM{{\cal M}}
\def\mpl{M_{\rm Pl}}
\newcommand{\bel}[1]{\be\label{#1}}
\def\al{\alpha}
\def\bt{\beta}
\def\eps{\epsilon}
\def\eg{{\it e.g.}}
\def\ie{{\it i.e.}}
\def\mn{{\mu\nu}}
\newcommand{\rep}[1]{{\bf #1}}
\def\be{\begin{equation}}
\def\ee{\end{equation}}
\def\bea{\begin{eqnarray}}
\def\eea{\end{eqnarray}}
\newcommand{\eref}[1]{(\ref{#1})}
\newcommand{\Eref}[1]{Eq.~(\ref{#1})}
\newcommand{\gsim}{ \mathop{}_{\textstyle \sim}^{\textstyle >} }
\newcommand{\lsim}{ \mathop{}_{\textstyle \sim}^{\textstyle <} }
\newcommand{\vev}[1]{ \left\langle {#1} \right\rangle }
\newcommand{\ev}{{\rm eV}}
\newcommand{\kev}{{\rm keV}}
\newcommand{\Mev}{{\rm MeV}}
\newcommand{\gev}{{\rm GeV}}
\newcommand{\tev}{{\rm TeV}}
\newcommand{\mev}{{\rm MeV}}
\newcommand{\mnu}{\ensuremath{m_\nu}}
\newcommand{\mlr}{\ensuremath{m_{lr}}}
\newcommand{\acc}{\ensuremath{{\cal A}}}
\newcommand{\mav}{MaVaNs}
\newcommand{\nusm}{$\nu$SM }

\begin{abstract}
We show  that for a force mediated by a vector particle coupled to a conserved $U(1)$ charge, the apparent range and strength can depend on the size and density of the source, and the proximity  to other sources. This ``chameleon''  effect is due to screening from a light charged scalar.  Such screening can weaken astrophysical constraints on new gauge bosons. 
As an example we consider the constraints on         chameleonic gauged $B-L$.     We show that although   Casimir   measurements greatly constrain any  $B-L\ $ force much stronger than gravity with range  longer than 0.1$\mu$m,  there remains an experimental window   for a long  range chameleonic  $B-L\ $ force. Such a force could be much stronger than gravity, and long or infinite range in vacuum,   but have an effective range near the surface of the earth   which is less than a micron.        \end{abstract}
\maketitle
\section{Introduction} 
Searches for equivalence principle violation  and deviations from the inverse square law   place stringent constraints on any new force. It is generally believed that new forces must be either very short range or much weaker than gravity. Since new $U(1)$ gauge interactions which are weaker than gravity are in conflict with known  string theory constructions, it is commonly assumed that any new gauge interaction must have range shorter than a few microns. However previous analyses of experimental constraints on new vector forces have  assumed linear equations of motion. While linearity is a reasonable approximation in many cases, it is never   exact \cite{cham}. Significant  nonlinearity can vastly weaken   the experimental constraints on gauge couplings for new forces. Nonlinear forces have been dubbed ``chameleons''\cite{chameleon}, and there have been several  analyses of constraints on scalar chameleons \cite{Steinhardt:2003iu,Mota:2003tc,Mota:2003tm,Gubser:2004uf,cham,Brax:2007vm,Olive:2007aj}. However  for new $U(1)$ gauge forces, all  previous analyses \cite{Adelberger:2003zx,Adelberger:2006dh} have assumed the forces   to be linear.      In this paper we will show that for  a new $U(1)$ gauge interaction, if the theory contains a charged scalar field,  then  the new force   is    chameleonic, and the constraints on its vacuum range and coupling strength can be substantially weakened. As a specific  interesting example, we consider the allowed parameter space for  a  chameleon  vector boson coupled to the difference between baryon and lepton number $(B-L)$.
\section{Quantum Gravity as Motivation for New Gauge Forces }
\label{intro}
Before launching into the study of experimental constraints, in this section we 
  consider  the theoretical  motivation  for   new long range forces, the motivation for chameleons, and review the theoretical  lower bound on the strength of the   coupling. 

There are many reasons in fundamental theory to consider new $U(1)$ forces, some of which have recently been summarized in an excellent review \cite{Langacker}. New forces are common features of string constructions, unified gauge theories,   and extra dimensional theories.
There has also long been interest in the explanation of the conservation laws for global charges, and speculation that new gauge invariances could play a role. Recently it has been argued that experimental detection of new forces could shed light on foundational  quantum gravity issues \cite{ArkaniHamed:2006dz}. 

A variety of quantum gravity based considerations   imply that there are no exact global symmetries in nature. For example, there are no global symmetries in string theory, an explicit candidate for a theory of quantum gravity. Also, global charges can disapear into black holes, which   then evaporate by emitting Hawking radiation, which  is independant of any global charge \cite{Bekenstein:1971hc}.  

Absence of exact global symmetry can be  reconciled with the experimental success of the Minimal  Standard Model (MSM). In the  MSM, Baryon number ($B$) and Lepton number ($L$) are anomalous, with only the combination $B-L$ preserved by anomalous electroweak processes.  At temperatures well below the weak scale, such anomalous processes occur exponentially rarely. Thus $B$ and $L$ are separately approximately conserved in the MSM, except   during    early times when the universe was extremely hot.  The conservation of $B$ and $L$ may be explained by the ``accidental'' absence of renormalizable gauge invariant   symmetry breaking operators involving the fields of the minimal standard model.   Provided that neutrino masses are Majorana, and provided the low energy particle states are simply those of the MSM,  it is possible that $B$ and $L$  conservation are merely an  inevitable feature of low energy physics, rather than any fundamental symmetry principle.  

Most extensions of the standard model do not share the feature of automatic accidental baryon and lepton number conservation. For instance if neutrinos are Dirac particles, then their right handed components are ``sterile'' under the standard model gauge interactions, and, unless forbidden by a new symmetry, a relevant lepton number violating dimension three operator 
 \bel{majorana2}
 M_{ij} N_i N_j + {\rm h.c.}
 \ee
 could give   Majorana masses to   $N_i$,  the three  right handed gauge singlets. Absence of such a term is evidence for a symmetry forbidding it. Thus Dirac neutrinos would be evidence for a new symmetry   which is not simply an accident of the emergent properties of the long range effective theory. Consistency with string theory and the previously discussed quantum gravity considerations would then suggest  a new gauge invariance principle.
 
 Even if neutrino masses are Majorana, there are compelling reasons to expect new gauge interactions.  
 There are many proposed extensions of the MSM,   motivated by a variety of considerations. The most compelling motivations are   the gauge hierarchy problem, and the strong evidence for dark matter. In most extensions of the MSM, such as  the Minimal Supersymmetric Standard Model (MSSM), the automatic nature of low energy $B$ and $L$ conservation is lost unless additional   symmetries are present. Usually it is assumed that the required additional symmetry is global, in apparent contradiction with quantum gravity, although it is possible in some models  for the additional symmetry to be a discrete remnant of a short distance gauge invariance. A new long range gauge interaction  could help explain why otherwise allowed terms are absent or suppressed. 
 
Naively, a new gauge invariance could be associated with an arbitrarily weak coupling. However according to  the Weak Gravity Conjecture  \cite{ArkaniHamed:2006dz}   any $U(1)$ gauge invariance must be   associated with a force   stronger than gravity.
  The argument derives from the holographic principle  \cite{'tHooft:1993gx,Susskind:1994vu,
Bousso:2002ju} which states that the amount of information which can be stored in any region is bounded by the boundary area in Planck units. This principle has been argued to be a likely feature of any theory of quantum gravity, and has been   used to argue that the number of types of stable black hole remnants must be finite.  
  In ref. \cite{ArkaniHamed:2006dz}, Arkani-Hamed et al. argue that the holographic principle implies that the  gauge coupling $g$ of any $U(1)$ gauge interaction must satisfy
 \be
 g> m_s/\mpl
 \ee
 where $\mpl$ is the Planck scale and $m_s$ is the mass of the lightest charged particle (here the charge is normalized to 1).
 In addition, they argue that  monopoles of mass less than $\mpl/g$ must exist. Since monopole masses are proportional to $\Lambda/g^2$ in an effective $U(1)$ gauge theory  with cutoff $\Lambda$, such a $U(1)$   must break down at or lower than a scale  
 \be\Lambda\sim g \mpl \ .\ee   
  Thus, for instance, if $B-L$ were   gauged, in an effective theory valid below 100 GeV, the weak gravity conjecture implies that the associated force between 2 neutrons from $B-L$ gauge boson exchange would have to be at least $\sim10^3$ times stronger than gravity, and the associated gauge coupling $g$ must satisfy
  \be g\gtap 10^{-17}\ .\ee 
 No such force has been seen by searches for equivalence principle or inverse square law violation.      
The $B-L$ force could be short range due to a Higgs mechanism from   a vacuum condensate of a $B-L$ charged scalar field. The phase of this field is eaten by the Higgs mechanism, but the modulus can be excited, implying the existence of a new scalar particle as well as a vector.
 
 In the next section we will show that such a boson implies a chameleonic nature for  the $B-L$ force, and so the  experimental bounds must be reexamined. We also consider  the case where there is  no vacuum condensate, but  there exists a light scalar with a positive mass which  condenses in matter and screens the $B-L$ force  for sufficiently large sources. We will argue that there is a parameter region such that the $B-L$ force is  substantially screened from detection near the surface of the earth. 

\section{Chameleon Vector Force}

In this section we show how a new $U(1)$ force can be weakened through screening by a scalar condensate.

As a specific example, we consider a renormalizable  extension of the Standard Model, in which  the $B-L$  symmetry is gauged. We call the new vector boson the ``paraphoton''.  To cancel gauge anomalies and allow  neutrino masses we add three right handed neutrinos. In addition, we add a complex scalar field $s$ which  carries charge $q$ charged under $B-L$, and has mass squared $m^2$. We will consider both possibilities for the sign of $m^2$.  

Ordinary   matter has net positive $B-L$ charge.  Since  a  force  proportional to $B-L$ has not been observed, either the coupling must be extremely weak, or there must be a mechanism to screen the charge. 

 When $m^2$ is positive,  in the vacuum there is no $s$ condensate, and the vacuum range of the $B-L$ force is infinite. We will show however that   any sufficiently large chunk of matter will contain an $s$ condensate and so the $B-L$ force between sufficiently large objects will  be short range. 
For negative $m^2$ there is a vacuum $s$ condensate and the $B-L$ force has finite range in vacuum. The chameleon effect must still be considered  in a constraint analysis, as   the effective range may be much shorter in matter.

The following argument demonstrates that  with a massless paraphoton,    it is energetically favorable  for any sufficiently large piece of ordinary matter to contain a condensate of  scalar $s$ particles to screen the $B-L$ charge. Note that due to Fermi blocking it is  energetically preferred to screen with light scalars than   with  light fermions such as neutrinos  \cite{Dolgov:1995hc}.   
We consider a sphere of constant charge density, and total $B-L$ charge $Q$. The total energy stored in the $B-L$ field per unit charge grows as  the square of the radius $R$.  For a sufficiently large object, the energy in the $B-L$ field  is large enough that it is energetically favorable to screen the field by forming  a condensate of $s$ bosons inside the object. The energy   per unit charge that would be stored in an unscreened field is
\beq\label{omegaeq}
\frac{E}{Q}&=&\frac{12  \pi g^2 Q}{5 R }= \frac{16 g^2   \pi^2 \rho R^2}{5 }  \cr 
&\sim&0.06 {\rm eV}\frac{ \rho  }{  10^{30} \rm m^{-3}}
\left(\frac{g}{ 10^{-8}}\right)^2\left(\frac{R}{ 10^{-5} \rm m}\right)^2 
\eeq
where $\rho$ is the neutron density and $E$ is the total field energy. Because the energy per unit charge grows with object size,  it is   favorable to screen the charge of sufficiently large objects.  In the following section  we will show that in the case where screening is energetically favorable, the scalar field effectively obtains a negative mass squared inside a charged object, hence an unscreened configuration is  unstable to $s$ condensate formation.

Since the energy per unit charge of a scalar condensate is  at least $ m/q$,  a light scalar is a prerequisite for  screening of the charge of small objects.
The smallest object separation for which sensitive force tests exist is $ \sim 10\mu$m, and typical neutron densities in ordinary solids are of order $10^{30}  \rm m^{-3}$.    For example,  with an $0.02 $eV mass scalar of $B-L$ charge $q\sim 1$, it is energetically favorable for the $B-L$ force exerted by a  20$\mu$m radius sphere to be screened,  provided $g\gsim 10^{-8}$. 
For nonspherical objects,  the object size $R$ in the above calculation should be replaced by a factor which depends on the shape as well as the overall volume of the object. 
For example, for a similar calculation for a charged disk the factor $R$ in the last line of \Eref{omegaeq} should be replaced with $c \sqrt{r R}$, where $c$ is a number of order 1, $R$ is the radius of the disk and $r$ is its thickness.  

 When $m^2$ is negative the paraphoton  acquires a vacuum  mass 
  \be
 m_V=| \sqrt2 q g \langle 0|s|0\rangle |\ .
 \ee
 In conventional interpretations of searches for new forces, a hypothetical  force is parametrized by a Yukawa potential of range $ m_V^{-1}$,  and strength $\alpha$  (between 2 neutrons) relative to gravity
 \be\label{conventional}
 \alpha\equiv \frac{g^2}{4\pi G_N m_N^2}\ .
 \ee
 
 However in matter, the value of the $s$ condensate can be substantially larger than it is in vacuum, reducing  the range of the vector force, and changing its apparent strength. Any massive  vector particle receiving mass from the Higgs mechanism has a chameleon nature.  However the chameleon  effect is   substantial only when  there exists a    scalar   whose potential is flat enough  so that the scalar expectation value  changes significantly in the presence of a background charge density.

 \subsection{  Approximations  for Chameleon Vector Force Calculations}   
 
 We now turn to more detailed  consideration of  the coupled $s$ and paraphoton dynamics, assuming an    $s$ condensate inside macroscopic chunks of ordinary matter. 
 
  We begin by reviewing scalar  chameleonic fields \cite{Steinhardt:2003iu,Mota:2003tc,Mota:2003tm,chameleon,Gubser:2004uf,cham,Brax:2007vm,Olive:2007aj}.
  
    For a real field $\psi$ coupled to a static source $j$, with potential $V(\psi)$, a static configuration which minimizes the total energy solves the equation of motion
\be
-\del^2\psi +\frac{\partial V}{\partial\psi } = j\ .
\ee
  For a constant source, a solution $\psi_0$ for $\psi$    can be found such that  
 \be
 \frac{\partial V}{\partial\psi }\bigg|_{\psi=\psi_0} = j\ .
 \ee

 The effective mass  for $\psi$ excitations in a constant background  field $\psi$ is 
 \beq
 m_{\rm eff}\ & =&\sqrt{ \frac{\partial^2 V}{\partial \psi^2}}\bigg|_{\psi=\psi_0}\ .   \eeq
 The  screening length   $  1/ m_{\rm eff}\ $ is the effective length scale  of the force mediated by the $\psi$ field inside a large constant density object. Effectively the field  is only sourced by the matter within this length scale.

 For an object of finite size which is much larger than $1/ m_{\rm eff}\ $,  
 physical arguments and numerical studies   \cite{cham} show that  $\psi\rightarrow \psi_0$ in the bulk of the object. Outside the surface of the object, $\psi$ falls rapidly on a scale of order $1/ m_{\rm eff}\ $.  Thus between large objects,   $1/ m_{\rm eff}\  $ is the effective range over which the  force is strong. Furthermore    only a ``thin shell'' of thickness $1/ m_{\rm eff}\ $ acts as a source for the field.  The vacuum range of the force is $1/m$, the inverse Compton wavelength of the particle associated with quantized vacuum excitations. When $1/ m_{\rm eff}\  $ is much shorter than  $1/m$, then,  on  scales between $1/ m_{\rm eff}\ $ and    $1/m$, there is a much weaker residual force. For the case of a quartic poential,  the   strength $\alpha_{\rm eff}\  $compared to gravity  of the residual force  is of order \cite{cham}
\be\alpha_{\rm eff}\  \sim \frac{1}{\epsilon M_1 M_2 G_N},\ee
where $\epsilon$ is the self-coupling \cite{cham}.

A significant chameleon effect for a gauge   particle   requires  at least  two  fields. We will consider a model with a charged scalar, $s$ and vector $B_\mu$, with matter acting as a source for the $B_\mu$ field, and the $B_\mu$ field in turn acting as  negative term in the potential  for $s$. 

The Lagrangian for our model is
  \beq\CL&=&(\partial_\mu+iqgB_\mu)s^*(\partial^\mu-i q gB^\mu)s-m^2|s|^2-\frac\epsilon2|s|^4\cr&&-
  \frac14F_{\mu\nu}F^{\mu\nu}-gB_\mu j^\mu \ .
  \eeq

Depending on the sign of $m^2$, this is either scalar QED or the Abelian Higgs model, with  a background charge source $j_\mu$. The scalar field $s$ will condense inside macroscopic objects and screen the charge.
To estimate the range of a chameleon force between macroscopic objects, it is useful to 
first  find the values of the $s$ and $B$ fields  deep inside the objects which will minimize the total energy. We can then determine  screening lengths for the $s$  and $B$ fields. 
 As for  a scalar chameleon, when both these screening lengths are short compared with the object size, a reasonable approximation  is to take the fields  inside the object to equal  the  values they would have inside an infinite sized object.
  Outside a   large object the fields drop to their vacuum values    over length scales of order their screening lengths.

We     consider  a  time independent background charge  charge density $j^0=\rho$, and $\vec{j}=0$.    To minimize the total energy we set the magnetic field to zero, and  allow an  $s$ field carrying a time independent charge density $\rho_s$
\be \rho_s= -i q( D_0 s s^*-s^* D_0 s)\ ,
\ee and no current density. We  work in a gauge with $\vec{B}=0$.
For a minimal energy configuration of given charge, we take $s$ to have the form $|s(\vec{x})| e^{-i q w t}$.    
  It is then convenient to define a  field \be\omega\equiv w +   g B_0\ ,\ee which is gauge invariant under the residual gauge invariance.
For a spatially uniform configuration,  $\omega$ is the energy per unit charge contained in the $s$ condensate, which, for  constant fields and $\epsilon=0$,   is simply the $s$ particle mass divided by its charge. In general, nonzero $\epsilon$ raises $q\omega$ to a value greater than $m$. We will refer to  $q\omega$ as the   inverse scalar screening length.   Note that  $\omega $  depends on the background charge density $\rho$.      
   
      To minimize the total energy for a static configuration where  $\omega $ and $|s|$ are position dependent, one must solve  the coupled, nonlinear equations of motion  
  \beq
\label{eom}
\del^2|s| &= &\big( m^2+  \epsilon |s|^2 - q^2\omega^2\big)|s| \cr
\del^2\omega  &=&- g^2\rho +2q^2g^2\omega |s|^2\ .
\eeq
 
Examination of these equations shows that  $\sqrt{2}qg|s| $ acts as the  effective mass for the gauge field, while the combination $ m^2+  \epsilon |s|^2 - q^2\omega^2$ acts as an effective mass squared for $s$. When this quantity is negative, it indicates an instability. 
 
 As a first step in finding an approximate solution for these fields in a macroscopic object,  
we  consider a spatially uniform background charge density $\rho$. We will refer to the  values of $\omega $ and $|s |$ which  minimize the total energy for uniform background charge $\rho$ as $\omega_0$ and $s_0$. Finite energy requires    the $s$ condensate to neutralize the charge density of the  background which implies
 \be\label{rhoeq}
-2 q^2 \omega_0 s_0^2=\rho_s= -\rho \ . 
\ee
The total energy density subject to this constraint is minimized when
\be\label{energyeq}
\omega_0^2= \frac{m^2+\epsilon s_0^2}{q^2} \  .
\ee
The equations \Eref{rhoeq} and \Eref{energyeq} allow us to solve for both $\omega_0$ and $s_0$ inside  large constant density objects, for either sign of $m^2$.
 
We can re-write \Eref{rhoeq} and \Eref{energyeq} as:
\be
s_0^2 = \frac{\rho}{2q}\frac{1}{q\omega_0} \qquad q^2\omega_0^2 = m^2 + \epsilon s_0^2
\ee
giving an equation for $q\omega_0$:
\be
2q\omega_0\Big(\big(q\omega_0\big)^2 - m^2\Big) - \frac{\epsilon\rho}{q} = 0
\ee
 
The solution is
\beq\label{wsol2}
\omega_0&=&\frac{1}{2^{2/3}q}\bigg[\bigg(\epsilon\rho/q + \sqrt{\big(\epsilon\rho/q\big)^2 - {\sfrac{16}{27}}m^6}\bigg)^{1/3} \cr
&&\quad\quad + \bigg(\epsilon\rho/q - \sqrt{\big(\epsilon\rho/q\big)^2 - {\sfrac{16}{27}}m^6}\bigg)^{1/3}\bigg]\cr
 \label{ssol}
s_0&=&\sqrt{\frac{\rho}{2 q^2 \omega_0} }\ ,\eeq which  gives manifestly real and nonnegative $s_0$ and $\omega_0$.
Because of the complexity of these formulae, it is convenient to consider three limiting cases where they simplify considerably.

\noindent {\bf Non Chameleon Case:}
For negative $m^2$, in the   limit 
\be\label{noeps} |m|  \gg (\epsilon\rho)^{\frac13} , \ee we
have
\beq\omega_0&\approx&\frac{\epsilon\rho}{2 q^2 \big|m^2\big|}\cr
&\quad&\cr
s_0&\approx& \sqrt{\big|m^2\big|/\epsilon}\ . \eeq 
The expectation value of the scalar field is approximately the same inside  a macroscopic object and in vacuum, and the chameleon effect is negligible.  The experimental constraints on this case have already been analyzed extensively \cite{Will:2001mx,Chiaverini:2002cb,Adelberger:2003zx,
Adelberger:2006dh}, and we will not consider them further.

\noindent {\bf Low Density Chameleon Case: }
For positive $m^2$, in the limit  of \Eref{noeps}, 
  \Eref{wsol2}  becomes
\beq\omega_0&\approx&m/q\cr
s_0&\approx&\sqrt{\frac {\rho}{ 2 q m}}\ .\eeq 
The charge density is low enough so that the parameter $\epsilon$ may be neglected, and the $s$ particle  self interaction is negligible.  In this case $m/q$ gives an approximate upper bound on the strength of  $\omega$, the static gauge field, and a scalar condensate will form in any sufficiently large charged object.  
There is a simple criterion for the minimal size.
The effective density dependent screening length of the paraphoton  inside a large   object  is given by 
\beq\label{veq}
\ell_V\equiv(\sqrt2 q g s_0)^{-1}&=& \sqrt{\frac{ \omega_0}{g^2\rho}}\cr
&\approx&  \sqrt{\frac{m}{q g^2\rho} }. \eeq 
 From  the definition \Eref{veq} of $\ell_V$, we see that the condition that $\omega_0$, the energy per unit charge  of a screened  configuration, should be  than the  unscreened energy per unit charge given in \Eref{omegaeq}, is  equivalent to the condition that the screening length is less than the size of the object,
\be\ell_V<R\ ,\ee   up to factors of order 1.

\noindent{\bf High Density Chameleon Case:} 
For either sign of $m^2$,  in the high density limit 
\be\label{condition2} |m|  \ll (\epsilon\rho)^{\frac13}\ee 
an approximate solution inside sufficiently large objects is
\beq\omega_0&\approx&\frac{(\epsilon\rho)^{\frac13}}{2^{\frac13}q^{\frac43}}\cr
s_0&\approx&  \frac{ \rho^{\frac13}}{\epsilon^{\frac16}  (2q)^{\frac13}}  \  . \eeq
   In this regime the  parameter $m^2$ is too small to be important.  The vector screening length is approximately\beq\label{veq2}
\ell_V \equiv(\sqrt2 q g s_0)^{-1}
&\approx& \frac{  \epsilon^{\frac16}}{g q^{\frac23}2^{\frac16}\rho^{\frac13}} .\eeq

\subsection{Chameleons and the thin shell condition} 
 
We now consider the properties of static fields for   charged objects of finite size $R$.  For    equations \Eref{rhoeq} and \Eref{energyeq} to yield the correct values of the vector and scalar fields deep inside the object, it is necessary that both  $\ell_{V,S}\ll R$, where $\ell_V$ is defined in \Eref{veq}, and   \beq\label{lscond}
 \lsm&\equiv&\frac{1}{\meff}=\frac{1}{\sqrt{m^2+\epsilon s_0^2}} \cr
&= &\frac{1}{q\omega_0}. \eeq
Then the total energy density for the field configuration is dominated by the volume inside the object, and so we expect the solution deep inside the objects to approach equation  \Eref{wsol2}.   We will refer to the case of $\ell_{S,V}  \ll R$ as the thin shell case.  From the definition of $\ell_V$ given in \Eref{veq}  and from the requirement that $\omega_0$ should be less than the unscreened $E/Q$ given in \Eref{omegaeq}, we see that $\ell_V\ll R$  is possible only if 
\be\label{gcondition}
g\gtap \sqrt{\frac1{\rho R^3}}\,
\ee
that is, the total charge of the object has to be much larger than $1/g^2$.

Throughout this paper, when estimating laboratory constraints on new forces, we will assume sources have a charge density of order $10^{30}/\rm m^3,$ a typical density for  solid matter.
For centimeter sized objects of typical solid density, a thin shell is possible only for $g\gtap 10^{-12}$. Far from the object, the fields fall to zero.   Examination of \Eref{eom} shows   that outside the object, in the region where $\omega\approx \omega_0$, that is, the region where the gauge field has not yet decayed,  the $s$ field does not decay either. Hence the scalar condensate always decays to its vacuum value at an equal or longer length scale than does the  gauge field. Note from \Eref{eom} that for constant $s$, the equation for $\omega$ is just a Yukawa equation.  Thus outside the object, the gauge field decays exponentially on a length scale $\ell_V$. Beyond $\ell_V$, $s$ decays with the density dependent  effective  length scale   $\lsm$.

 \subsection{Chameleons Beyond the Thin Shell Condition}
 \label{beyond}
 We now consider the case  of smaller objects.
If $\ell_V>R$, the entire object sources the gauge field. The total energy stored in the gauge field is then not substantially affected by screening. It is therefore energetically favorable to have the scalar field take its vacuum  value, and there is no significant chameleon effect.  There could however be a large chameleon effect with   $\ell_S>R$.   In this case the fields do not approach constant values inside the object and we must rely on numerical analysis for an  accurate approximate   solution. For order of magnitude estimates, we may use energetic  and dimensional analysis considerations, as we do here.

If we neglect the scalar potential, we may  easily obtain an   estimate of the  order of magnitude of the scalar and vector fields inside a spherical  object of radius $R$. It is reasonable to assume that the value of $s$ does not vary by more than an order of magnitude inside the object in this regime, since there is no parameter to set a shorter length scale for  the variation of  $s$. We call the value of $s$ at the surface of the object $s_0$.  We assume \be\label{sbound}s_0\gg \langle 0|s|0\rangle\ ,\ee the vacuum expectation value of $s$, since otherwise there is no significant chameleon effect. We   then neglect $\langle 0|s|0\rangle$. Assuming the energy stored in the $s$ field is dominated by kinetic  rather than potential terms,  this energy scales as $s_0^2$, and by dimensional analysis is of order $s_0^2 R$. The screening length of the vector field  in the vicinity of the object is of order $ 1/|g  q s_0|$.  For $|R g q s_0|\ll1$, the total energy stored in the vector  field is of order $g^2 \rho^2 R^3/(g^2 q^2 s_0^2)$. Minimizing the combined energy leads to the order of magnitude estimate for the surface value of $s$ in this regime
\be s_0\sim \sqrt{\frac{\rho R}{q}}\ .\ee Note that this estimate is independent of $g$, but does assume  
that  \Eref{gcondition} is satisfied.

Comparison of the value of  the scalar field kinetic energy and potential energy shows that our neglect of the scalar potential is reasonable as long as  $m^2\ll 1/R^2$ and $\epsilon\ll1/(\rho R^3)$. For larger values of $m$ or $\epsilon$  the scalar potential is relevant.  For either $m \gg 1/R $ or $\epsilon\gg 1/(\rho R^3)$  we may use the thin shell approximation discussed in the previous section.

 \section{Experimental Constraints on Chameleonic $B-L$}
 \subsection{Laboratory constraints on non thin shelled chameleons}
If we  use the results of the previous section to examine the force between two similar non thin shelled objects, we must take into account that although the vector fields fall off very rapidly over a scale much shorter than $R$, the scalar fields do not, and will mediate an attractive force between objects with a range $\tilde\ell $ of order $\lsm$ which we  assume  to be longer than the typical object size. We may estimate the strength of such a force by comparing the value of $s_0$ with $x$, which is defined to be the value
 a massless scalar field which was sourced by a Yukawa coupling $y_{\rm eff}\ $ would have at the surface, 
 \be 
 x=\frac{y_{\rm eff}\ \rho R^2 }{\sqrt{4\pi}} ,\ee 
 to find an effective Yukawa coupling
 \be y_{\rm eff}\ \sim\sqrt{\frac{4\pi}{q\rho R^3}} . \ee  
 Thus the force between two similar objects would have strength  compared with gravity $\alpha_{\rm eff}\ $ which is roughly of order
 \be\label{strength}
 \alpha_{\rm eff}\ \sim \frac{y_{\rm eff}\ ^2}{4 \pi G_N m_N^2}\sim\frac{1}{q m_n^2 G_N \rho R^3}.\ee For $q\sim 1$, this force is weaker than gravity only  for objects with total mass greater than  about $10^{38}$ GeV which is about  $10^{11}$ kg.
 
 These estimates are very rough, and have made oversimplifying assumptions. However   the estimates imply  that laboratory sized   objects with  a non thin shelled scalar condensate attract each other via a new force   which is many orders of magnitude   stronger than gravity. It is clear that  unless we consider light charged scalars with exponentially large values of $q$, table top experiments rule out any parameter  regime where laboratory sized objects have their charges screened by a non thinshelled scalar condensate.

\subsection{Casimir Constraints}
 Ordinary materials with an $s$ condensate are very good $B-L$ charge conductors. In any $m^2>0$   thin shelled chameleon case, gauged $B-L$ should result in a Casimir force   from $B-L$ quantum fluctuations, in addition to the Casimir force for electromagnetism \cite{Casimir:1948dh}. Similarly,  even with  $m^2<0$, if the vacuum screening length is $\lv$,  there is a Casimir effect at distance scales  between   $\lv$ and $\lvm$. In the traditional Casimir effect  \cite{Casimir:1948dh,Milton:2002vm} between perfectly conducting parallel plates, the free electrons enforce boundary conditions on the photon, limiting the available modes between the conductors and creating an inward pressure on the conductors.  This attractive force is completely independent of the  gauge coupling for perfect conductors, and the gauge coupling only becomes relevant in the regime of finite conductivity when the skin depth of the conductor is well above the plate separation, destroying the boundary conditions imposed on the vector field.

In a perfect conductor, the free electrons will impose the same boundary conditions on the paraphoton as the photon.   This Casimir force due to the existence of the paraphoton equals the Casimir force due to the photon, seemingly far outside of experimental constraints.  But the gauge coupling affects the magnitude of the Casimir force in the case of finite conductivity, which is the relevant case for experimental constraints.

Modern experiments have determined the Casimir force within 5\% of its theoretically predicted value over a range of geometry scales of nearly two orders of magnitude.  If   $B-L$ gauge bosons are light, then the Casimir force due to $B-L$ must be suppressed by some mechanism.  Since the relevant gauge boson couplings are to electrons for the Casimir effect, the scaled gauge coupling provides the means to suppress the $B-L$ Casimir force.  For materials with finite conductivity, the skin depth can exceed the plate separation,   relaxing the boundary conditions imposed by the conductors and suppressing the Casimir force.  

In practice, to reduce the Casimir force beyond an observable level, the skin depth must greatly exceed the plate separation.  Lamoreaux \cite{Lamoreaux:1999cu} has calculated corrections to the Casimir force between parallel plates for three real materials relevant to experiments and found the correction factor is above 0.6 when the skin depth is on the order of the plate separation.  Therefore, we do not expect the $B-L$ Casimir force to subside until the skin depth is much larger than the plate separation.  The skin depth is related to the plasma frequency in the metal by
\be
\delta = \frac{1}{\omega_p}
\ee
where the electromagnetic plasma frequency $\omega_p$ is
\be\label{omegapeq1}
\omega_e = \bigg(\frac{  \rho_e e^2}{m_e}\bigg)^{1/2}
\ee
and $\rho_e$ is the free electron density, and $m_e$ the electron mass. 
For the $B-L$ force the $s$ also contributes to the conductivity. Replacing $\rho_e$ in   \Eref{omegapeq1} 
by the scalar charge density $ q \omega_0 |s|^2$, $e$ by the $B-L$ gauge coupling $g$, and $m_e$ by  the effective mass/unit charge $\omega_0$,  gives the paraphoton  plasma frequency $\omega_p$ to be  
\be\label{omegapeq2}
\omega_p = m_V\ .
\ee
and
\be
\delta =   \ell_V  \ .
\ee
Current tests span the separation range of 0.1 to 6 $\mu$m \cite{Lamoreaux:1996wh,Lamoreaux:1996wh,Mohideen:1998iz,Roy:1999dx,Bordag:2001qi}, with the best bound coming in the earliest modern Casimir force measurement by Lamoreaux \cite{Lamoreaux:1996wh}, which used Au coated plates with a 6$\mu$m separation, and thickness of 0.5 cm. 

As discussed in the next section, as long as $q\sim 1$ we can safely assume that $\lsm$ is less than the plate separation or else the energy contained in the scalar field would imply a large force between the plates. 
Thus a new $B-L$ force  with vacuum range $\lv > 0.1\ \mu$m,   and screening length $\lvm\ll\lv$   is ruled out by the paraphoton quantum fluctuation contribution to   Casimir tests. 
This would seem to constrain the vacuum range of any new chameleon gauge force much stronger than gravity to be less than $\sim0.1\ \mu$m. That is to say, the vacuum mass of any new gauge boson with a coupling strong enough to lead to a chameleon effect in laboratory objects must be greater than of order 
 2 eV. 
However, we note that all Casimir measurements are performed near the surface of the earth. Since the earth is a very large object, we must consider its effect on the values of the fields near its surface. We will address this effect in  \S \ref{earth}.

   \subsection{Short range tests of the   Inverse Square Law}

   These tests will provide  highly complementary constraints to the Casimir constraints considered in the previous section as they give strong upper bounds on $g$ as a function of  $\lvm$ and $\lsm$. 
   
We assume the chameleon force is largely screened  at distance scales larger than the greater of $\lvm$ and $\lsm$.

  In the $\lsm<\lvm$ limit, we have a repulsive force   with constraints are similar to those on a force of relative strength given by \Eref{conventional}, and effective range $\lvm$. 
  
In the previous section we saw that Casimir constraints require $\lvm$ to be larger than 6 $\mu$m, unless $\lv$ is shorter than $0.1\mu$m or  unless $\lsm > 6 \mu$m.
However   searches for new forces  at short distances have put constraints on new forces in this range. The strongest   constraints at 6 $\mu$m come from  an   experiment  at Stanford \cite{Chiaverini:2002cb}. Any   force in this range must have $\alpha_{\rm eff}\ < 10^9$   or  $g < 10^{-14}$. According to \Eref{gcondition}, for $q\sim 1$, this value of $g$ is too small to allow for an $s$ condensate in a metallic density  object of size less than  of order 1000 km. Similarly,   when $\lsm$ is longer than 6 $\mu$m, there will be   additional energy density in the $s$ condensate   which will lead to an attractive force between the objects. An estimate of the force per unit area between parallel plates of separation less than $\ell_S$ is 
just the energy density in the condensate which is
 \be F_s/A=\rho \omega=\frac{\rho}{ q \ell_S} \ .\ee 
 We may compare this with the gravitational force/unit area $F_g/A$ between two parallel plates  of thicknesses $L_1$ and $L_2$ and mass density $\rho_m$   
  \be F_g/A=\frac{\rho_m^2 L_1 L_2 G_N}2\ .\ee
 For thin plates the ratio $\alpha_{\rm eff}\ $ between these two forces is enormous  
 \beq\label{scalarforce}&&\alpha_{\rm eff}\  \approx  2\times 10^{14}\cr&&\times\left(\frac {2 (\rho/10^{30}{\rm m}^{-3}) }{q (\rho_m/( 2 {\rm gm\  cm}^{-3}))^2 (L_1/1 {\rm cm})(L_2/ 1{\rm cm})(\ell_S/ 1{\rm cm})} \right). \nn\eeq
Clearly, for $q\sim 1$, the case $\lsm>\lvm$ is ruled out for a thin shelled chameleon. As we saw in \S \ref{beyond}, the force between non thin shelled chameleons is also very strong compared to gravity.   Thus the results on searches for new forces in the $6 \mu$m range, when combined with the Casimir measurements, and the lower bound on the gauge coupling for a chameleon effect, would seem to rule out any  chameleon force  with vacuum range longer than $ \sim0.1\ \mu$m.   A possible loophole, discussed in the next section,  is that the these experiments are not done in isolation.

\subsection{The effect of doing experiments near a large source: the earth}
\label{earth} 
The most stringent searches for new forces either use the earth as a source or take place near the surface of the earth.

The total $B-L$ charge of the earth $Q$ is of order 
\be
Q\sim  10^{51}\ . \ee 
We see from \Eref{gcondition}  that provided there exists a  light scalar  of charge  of order one,   the lower bound on the coupling  for chameleonic screening  of the charge of the earth is
\be g \gtap 10^{-25}\ .   \ee   Thus any force strong enough to be seen  in any current experiment can be screened in the earth.  

An interesting chameleon region is where  $\lvm$ is shorter than the laboratory limit of $\sim 0.1 \mu m$ but   $\lsm$  is much longer. Then  no laboratory sized object will act as a significant chameleon source within a distance  from the surface of the earth of order $\lsm$. A force  with {\it vacuum} range  of infinity   could escape detection  in any experiment  which either uses the earth as a source or is closer to the  surface of the earth than $\lsm$, unless the experiment is sensitive to a distance shorter than $\lvm$. A scale $\lsm $ of order  a few hundred meters is  sufficient  to eliminate the inverse square law violation  constraints on forces with   $\lvm<0.1\mu$m, independent of the vacuum range. 

\subsection{Chameleons and Long Range Gravitational Tests}

The strongest constraints on the strength of new forces arise at  ranges longer than 10 km,  from tests of the inverse square law and equivalence principle violation. In ref. \cite{Will:2001mx,Adelberger:2003zx,Adelberger:2006dh} a variety of constraints are given. In the absence of screening, any new force of range longer than $\sim 10^7$ m,  coupled to $B-L$ must have coupling  $g$ less than   $10^{-24}$ from equivalence principle violation searches. This limit is slightly stronger for a force of range comparable  to the earth-moon distance.  However, we can use \Eref{omegaeq} to show that  for a   long range $B-L$ force with coupling greater than $10^{-24}$, provided there exists a charged scalar lighter than $10^{-9} $eV, the $B-L$ charge of the earth will be screened and the force will not be detectable in experiments using the earth as a source.   

In ref. \cite{cham} it was argued that measurements of $G_N$ inferred from measurements of $g$ at various lake ocean depths    constrain  new forces which are unscreened in laboratory Cavendish measurements  but are screened in the earth to be less than  $  10^{-3}$ gravitational strength, corresponding to $g\sim 10^{-20}$.  These limits will not apply if $\lsm$ is of order a hundred meters, corresponding to a scalar lighter than $\sim 10^{-9}$ eV.

We conclude that a new $B-L$ force with astronomical range  in vacuum may  be screened in any searches for new forces using large sources such as the earth, or taking place near the surface of the earth.   The main constraint on the strength of such a force would come from precision particle physics measurements, which, as discussed in \S \ref{atom} and ref. \cite{us}, are sensitive to couplings as small as $g\sim 10^{-5}$.

\subsection{Chameleon forces and astrophysics}

Any new particle which is lighter than a few MeV and long mean free path in matter, must face stringent bounds on its couplings from supernovae and stellar cooling \cite{raffelt}. Such a particle could be emitted from the bulk of a star or supernova and result  in higher thermal conductivity or energy loss than is consistent with astrophysical constraints.  As discussed in  ref. \cite{us},  such bounds can be avoided for chameleon vector bosons, whose masses scale as the cube root of the density in the high density regime, allowing for Boltzmann suppression of the effects in stellar cores and supernovae. Thus new chameleon forces can have a range and strength which makes them   relevant for neutrino physics \cite{us} or for dark matter annihilation  \cite{Fayet:1980rr,Fayet:1990wx,fayet}.

\subsection{ Atomic Physics Constraints}
\label{atom}
 Constraints on short range forces which are stronger than gravity can come from precision atomic physics measurements. For instance in the case of gauged $B-L$, 
the presence of the paraphoton alters the Coulomb potential, an effect that is most easily realized in energy level shifts in single electron atoms.  Except for neutrons, the paraphoton and photon couple to matter in a similar way, although with different strengths and range.   When the range of the paraphoton is much longer than  the Bohr radius of an atom, then we may approximate the potential as Coulombic.

The Coulomb potential from a nucleus with atomic number $Z$ and mass number $A$ is
\be
V(r) = -\frac{Ze^2 + Ag^2}{r}
\ee
In single electron atoms, this gives rise to the energy levels
\be
E_n(Z,A) = \frac{m_e}{2\hbar^2n^2}(Ze^2 + Ag^2)^2
\ee
where we just want the leading order correction due to the paraphoton.  It is useful to put the energy levels in terms of the Rydberg, since it is very precisely measured.  In the presence of the paraphoton, the Rydberg is
\be
\textrm{Ry} = \frac{m_e}{2\hbar^2}(e^2 + g^2)^2 = \frac12m_e\alpha_e^2
\ee
where any mass number dependence in the energy levels will appear for higher $A$ atoms.  Note that the measured value of $\alpha_e$ would be $4\pi(e^2 + g^2)$, since fine structure constant measurements almost exclusively take place in systems where the paraphoton and photon are indistinguishable.  Then, with $N = A - Z$,
\be
E_n(Z,A) = \frac{\textrm{Ry}}{n^2}\Big(Z + \frac{g^2}{4\pi\alpha_e}N\Big)^2
\ee
Therefore, the leading order correction to the energy levels is given by
\be
E_n(Z,A) - E_n(Z,A,g=0) = \frac{2ZN}{n^2}\textrm{Ry}\frac{g^2}{4\pi\alpha_e} + \mathcal{O}(g^4)
\ee
Since this correction term must be less than the discrepancy between the currently best understood experimental and theoretical values of the energy levels,
\be
g^2 < \frac{n^2}{ZN}\frac{\delta E_n^{\textrm{exp-th}}(Z,A)}{\textrm{Ry}}2\pi\alpha_e
\ee
It is clear the best bound comes from ground state energy levels -- the bound is reduced by a factor of $n^2$ over other levels, and the theoretical and experimental values are much more precisely determined in general.  In fact, since the absolute energy of the level must be measured, the experimental errors will greatly exceed the theoretical ones, giving the bound for these levels of
\be
g < \sqrt{\frac{2\pi\alpha_e}{ZN}}\Big(\frac{\delta E^{\exp}_1(Z,A)}{\textrm{Ry}}\Big)^{1/2}
\ee
Since it is generally difficult to create high Z single electron atoms, a simple bound comes from $^4$He.  If $\delta E^{\exp}_1(2,4) = 10^{-5}$ eV, then
\be
g < 6.5\cdot10^{-5}
\ee 
Along with this constraint, there are constraints at all orders in $\alpha_e$, from the fine structure to the Lamb shift.  These constraints arise from two distinct effects, the mass number effect described above and new vacuum polarization effects on the paraphoton, arising from interactions of the paraphoton with the $s$.

A well known constraint for new light gauge bosons comes from measurement of the anomalous magnetic moment of the electron \cite{Gabrielse}, which requires \cite{us} that for a new boson which is much lighter than the electron,  $g< 1.7 \cdot 10^{-5}$.
\subsection{Collider Tests}
 
For completeness, we note that   couplings for  heavier paraphoton masses  are constrained   by collider experiments, including precision electroweak measurements. The existence of the $s$ field will not weaken these constraints.   In principle   $e^+e^-$ scattering cross sections set a limit on  the couplings of new forces of any range longer than of order $10^{-18}$ m.  The cross section for paraphoton radiation would be suppressed over photon radiation by a factor of $g^2/e^2$, and the paraphoton would only affect the total cross section $e^+e^- \rightarrow X$ by a factor of $g^4/e^4$.   Precision electroweak corrections and the effects of mixing with the $Z$ boson are also suppressed by $g$. If $g \lsim 10^{-2}$, then LEP and other $e^+e^-$ experiments would not have detected a  paraphoton. 

 \subsection{Cosmological Constraints}
Stringent cosmological constraints on new light particles come from nucleosynthesis, and from the observed fluctuations in the Cosmic Microwave Background (CMB). Current abundances of Helium, Deuterium, He$^3$, and Li$^{6,7}$ are fairly consistent with a standard model calculation in which at most   1 additional  light species of neutrino is in thermal equilibrium. The calculated Helium abundance, already a bit high compared to observation in the Minimal Standard Model, would come out even higher if   either  the paraphoton or the charged scalar were lighter than an MeV and had a thermal abundance \cite{Dolgov:2002wy,Chu:2006ua,Strumia:2006db,Hannestad:2006zg,Steigman:2007xt}.  A possible loophole occurs if the scalar potential is flat enough. Then there could be a large scalar condensate during the nucleosynthesis epoch, which would make the paraphoton much heavier than the temperature, and reduce the cross section for producing $s$ particles. Alternatively, the  paraphoton coupling  could be  small enough  so that paraphoton and $s$ production rates are not in thermal equilibrium during nucleosynthesis. In the absence of a large $s$ condensate, the most severe constraint  comes from comparing the rate for  paraphoton radiation  with the expansion rate of the universe. Paraphoton production  is of order $\alpha_{\rm em} ^2g^2 T$, where $T$ is the nucleosynthesis temperature of about an MeV,  $\alpha_{\rm em} $ is the fine structure constant, and the expansion rate of the universe scales as $T^2/m_{\rm Pl}$. Thus  the paraphoton can remain out of equilibrium provided that  $g\lsim \sqrt{T/m_{\rm Pl}}/\alpha_{\rm em} \sim 10^{-9}.$ If the paraphoton  is in equilibrium it constributes   slightly more than an additional neutrino species to the total energy density, which is marginally disfavored. Similarly,  if the  right handed neutrinos and the $s$ particle are   in equilbrium this would be problematic.  Since these production rates go as $g^4 T$, 
\be
g\lsim 10^{-6}
\ee
 is small enough. 
 
 In addition, there are CMB constraints on the number of new light species in equilibrium  around the time of recombination, and on the  scattering cross sections of neutrinos  \cite{Cirelli:2006kt,Mangano:2006ur,Hamann:2007pi,Hannestad:2007dd}, although currently these limits do not   constrain the  parameters as much as  nucleosysnthesis.

\section{ A Model for Exponentially Large Scalar Charge}
In the previous sections we have assumed that  $s$ carries charge of order 1, although a scalar carrying exponentially large charge would be effective at allowing a new force to evade existing constraints via screening. In this section we   discuss
 a possible mechanism to get an  new force which is apparently weaker than gravity  in the visible sector, although stronger in a hidden sector.  Consider a new $U(1)$ force operating in a hidden sector with a long but finite range $\ell_{\rm hidden}$, with coupling constant $g'$,  and  a new, shorter range $U(1)$ force  in the visible sector, with range $\ell_{\rm visible}$, and coupling constant $g$.   Kinetic mixing between the two new $U(1)$ forces could lead the longer range hidden sector force to couple to the visible sector with an effective  coupling to visible sector fields as large as $g \ell_{\rm visible}^2/\ell_{\rm hidden}^2$. Note that the coupling $g'$ to a hidden sector field could be much larger. This model could also provide a mechanism for large apparent $q$. If $s$ is part of the hidden sector this could provide a mechanism for $q$ as large as  of order $ (g' \ell_{\rm hidden}^2)/( g \ell_{\rm visible}^2)$, allowing for  a   viable chameleon force.

 \section{Conclusions}

We have shown that new vector forces can exhibit similar chameleon effects as scalar forces, that is, the effective range of the force can vary according to the size and density of the source. This variation is substantial when there exists a  light scalar  which carries the new force charge.   This scalar  will condense inside macroscopic objects and screen the force. For the case of infinite vacuum range,   the chameleon effect will be significant in any sufficiently large source. For forces with finite vacuum range    the chameleon effect is important in any  large source of sufficiently high density.

We have reconsidered the limits on gauged $B-L$, in the presence of a light scalar. We find that    the chameleon effect 
can greatly weaken the constraints from searches for inverse square law violation or equivalence principle violation. However such a chameleon vector boson can contribute to the Casimir force, and when Casimir constraints are considered,    a new long range force     should be be much weaker than gravity with two possible loopholes: either the charge carried by the scalar field is exponentially large,  or the scalar is  lighter than  of order $10^{-9}$ eV. Similar  conclusions   apply to gauged Baryon or Lepton numbers.

There are several possible interesting   allowed regions for a vector chameleon. One interesting case  is a chameleonic vector force which is   screened for any object larger than a few kilometers in size, or near the surface of such an object.  Detection of such a boson would require  both the source and detector to be small, and  located in space. Another interesting chameleon is  a vector boson which locally is light enough and strongly enough coupled to affect particle physics experiments, but which is much heavier in extreme   environments such as the cores of stars and supernovae. 
Applications for the latter chameleon   include   possible effects on neutrino oscillation experiments \cite{us} and on dark matter annihilation  \cite{Fayet:1980rr,Fayet:1990wx,fayet}.

 \section*{Acknowledgments} We would like to acknowledge useful conversations with Neal Weiner. This work  was partially
supported by the DOE under contract DE-FGO3-96-ER4095.  The work of Jonathan Walsh was supported by a LHC Theory Initiative Graduate Fellowship.  Ann Nelson would like to acknowledge partial sabbatical support from the Ministerio de Educaci\'on y Ciencia of Spain, partial sabbatical support from the University of Washington,  and the hospitality of the Insituto de F\'isica Te\'orica UAM/CSIC at the Universidad Aut\'onoma de Madrid.
 
\bibliographystyle{apsrev}

\end{document}